\documentstyle[pre,aps]{revtex}

\begin{document}

\draft

\title{Self--Similar Interpolation in Quantum Mechanics}
\author{V. I. Yukalov$^{1,2}$, E. P. Yukalova$^{1,2}$, and S. Gluzman$^{3}$}
\address{$^1$Centre for Interdisciplinary Studies in Chemical Physics \\
University of Western Ontario, London, Ontario N6A 3K7, Canada \\
$^{2}$Bogolubov Laboratory of Theoretical Physics\\
Joint Institute for Nuclear Research, Dubna, Russia \\
$^{3}$International Centre of Condensed Matter Physics \\
University of Brasilia, CP 04513, Brasilia, DF 70919--970, Brazil}
\maketitle

\begin{abstract}
An approach is developed for constructing simple analytical formulae
accurately approximating solutions to eigenvalue problems of quantum
mechanics. This approach is based on self--similar approximation theory. In
order to derive interpolation formulae valid in the whole range of
parameters of considered physical quantities, the self--similar
renormalization procedure is complimented here by boundary conditions which
define control functions guaranteeing correct asymptotic behaviour in the
vicinity of boundary points. To emphasize the generality of the approach, it
is illustrated by different problems that are typical for quantum mechanics,
such as anharmonic oscillators, double--well potentials, and quasiresonance
models with quasistationary states. In addition, the nonlinear Schr\"odinger
equation is considered, for which both eigenvalues and wave functions are
constructed.
\end{abstract}

\vspace{1cm}


\section{Introduction}

A standard problem in quantum mechanics is how to solve approximately
stationary Schr\"{o}dinger equations that do not possess exact solutions. In
the cases of asymptotically small and asymptotically large coupling
parameters, one may employ perturbation theory presenting solutions as power
series in powers of the corresponding small parameter. However, such series
are practically always only asymptotic and quickly diverge for sufficiently
small expansion parameters. Moreover, physical quantities of interest
usually correspond neither to weak--coupling nor to strong--coupling limits,
but to an intermediate region of a coupling parameter. Thus, the problem
arises of how to construct an interpolation formula, valid in the whole
region of physical variables, when only asymptotic expansions near
boundaries are known.

The most known method of deriving interpolation formulae is the two--point
Pad\'{e} approximation [1-3]. In some cases the latter yields quite
reasonable results. Nevertheless, the usage of this method has not become
widespread because of the following difficulties:

First of all, to reach sufficient accuracy by employing Pad\'{e}
approximants, one needs to have tens of terms in perturbative expansions
[1-3]. But the standard situation in physically interesting problems is when
one has in hands only a few terms. In such a case, for the same problem one
may construct different two--point Pad\'{e} approximants, all having correct
left--side and right--side limits, but differing from each other in the
intermediate region by $1000\%$ [4]. This clearly shows that in the case of
short series the two--point Pad\'{e} approximants cannot provide even
qualitative description.

Second, two--point Pad\'{e} approximants can treat at infinity only rational
powers [1-3] and are not able to describe other types of behaviour, for
example, power laws with irrational powers or exponential functions.
However, more complicated than the rational--power behaviour often occurs in
physical problems. For instance, exponential behaviour at infinity is
constantly exhibited by wave functions.

Third, interpolation between two different expansions, by using two--point
Pad\'{e} approximants, can be accomplished solely, when these two expansions
have compatible variables [1-3]. For example, even for a such a simple
problem as the anharmonic oscillator, the eigenvalues in the weak--coupling
and strong--coupling expansions have incompatible variables [5].

Fourth, there exists the well--known and annoying problem of appearance of
poles in Pad\'{e} approximants, which results in unphysical singularities
[1-3]. Eliminating such singularities in two--point Pad\'{e} approximants is
often impossible because of restrictions that are imposed by prescribed
boundary conditions.

Finally, Pad\'{e} approximation is rather a numerical technique, but we keep
in mind {\it analytical }approach that would combine relatively simple
representation for physical quantities with their good accuracy. The
advantage of having analytical expressions, as compared to just numbers
obtained from a numerical procedure, is in the convenience of analyzing such
expressions with respect to physical parameters entering into them.

In the present paper we develop an analytical approach for deriving
interpolation formulas, which is free of the above deficiencies of Pad\'{e}
approximation. This approach works well when just a few terms of asymptotic
expansions are available; it successfully sews power--law with exponential
asymptotic behaviour; it does not have at all the problem of compatibility;
no unphysical poles arise; it combines analytical representation with good
accuracy. We illustrate the approach by several quantum--mechanical problems
that are usually considered as typical touchstones for any new method. These
problems include calculation of energy levels for different anharmonic
oscillators, for the Hamiltonians with double--well potentials, and for
quasiresonance models. Each of these problems has its own specific
calculational difficulties (for review see Refs.[5-7]). This is why it is
important to show that all of them can be treated by the same approach.
Moreover, we demonstrate that the same method is applicable to the {\it %
nonlinear} Schr\"{o}dinger equation, for which we find both the energy and
wave function of the ground state. The latter example is interesting not
only as an illustration of wide possibilities of the approach, but it is
important for practical purpose, being related to the description of
Bose--condensed particles in traps. We carefully compare the properties of
the wave function we have derived with those of the Thomas--Fermi and
variational--Gaussian approximations. The analysis proves that our wave
function provides the best approximation for a solution to the nonlinear
Schr\"{o}dinger equation considered. Analytical expressions for the wave
function of a vertex filament are also constructed, being in good agreement
with numerical data.

\section{Self--Similar Interpolation}

Assume that we are interested in finding a function $f(x)$ in the interval 
$x_{1}$ $\leq x\leq x_{2}$. The latter can be finite or infinite. Let
equations defining the function $f(x)$ be rigorously unsolvable, so that
only perturbative asymptotic expansions near the boundaries can be derived:
near the left boundary, 
\begin{equation}
f(x)\simeq p_{k}(x,x_{1}),\qquad x\rightarrow x_{1}+0,
\end{equation}
and near the right boundary, 
\begin{equation}
f(x)\simeq p_{k}(x,x_{2}),\qquad x\rightarrow x_{2}-0,
\end{equation}
where $k=0,1,2,...$ For the time being, we do not specify the physical
nature of the function $f(x)$ and its variable $x$, since the general scheme
does not depend on these specifications.

In this section we develop such a general scheme for constructing
approximations to the function $f(x)$, so that these approximations,
interpolating between the asymptotic expansions (1) and (2), could be valid
in the whole region $x_{1}$ $\leq x\leq x_{2}$. The approach we develop is
based on the self--similar approximation theory [8-13] in its algebraically
invariant formulation [14-16]. Here we show how to construct self--similar
approximations so that they be compatible with the asymptotic boundary
conditions (1) and (2). Since all theoretical foundation and basic technical
details of the method have been expounded in our previous papers [8-16], we
do not repeat them here but only delineate the scheme of the approach
adapting it to the considered problem of interpolation.

Let us take an asymptotic expansion, like (1) or (2), in the vicinity of a
point $x_{i}$, with $i=1,2$. Define the algebraic transform 
\begin{equation}
P_{k}(x,s,x_{i})=x^{s}p_{k}(x,x_{i}),
\end{equation}
whose inverse, evidently, is 
\begin{equation}
p_{k}(x,x_{i})=x^{-s}P_{k}(x,s,x_{i}).
\end{equation}
Introduce an expansion function $x(\varphi ,s,x_{i})$ by means of the
equation 
\begin{equation}
P_{0}(x,s,x_{i})=\varphi ,\qquad x=x(\varphi ,s,x_{i}).
\end{equation}
Substituting this expansion function into Eq. (3), we obtain 
\begin{equation}
y_{k}(\varphi ,s,x_{i})=P_{k}(x(\varphi ,s,x_{i}),s,x_{i}).
\end{equation}
Transformation inverse to (6) reads 
\begin{equation}
P_{k}(x,s,x_{i})=y_{k}(P_{0}(x,s,x_{i}),s,x_{i}).
\end{equation}
The family $\{y_{k}\}$ of the endomorphisms defined in (6) is called [11-13]
the approximation cascade, because its trajectory $\{y_{k}(\varphi
,s,x_{i})\}$ is bijective to the sequence of approximations $%
\{P_{k}(x,s,x_{i})\}$. The cascade velocity can be given by the finite
difference 
\begin{equation}
v_{k}(\varphi ,s,x_{i})=y_{k}(\varphi ,s,x_{i})-y_{k-1}(\varphi ,s,x_{i}).
\end{equation}
The evolution equation, written in the integral form is 
\begin{equation}
\int_{P_{k-1}}^{P_{k}^{*}}\frac{d\varphi }{v_{k}(\varphi ,s,x_{i})}=\tau ,
\end{equation}
where $P_{k}=P_{k}(x,s,x_{i})$, the upper limit $P_{k}^{*}=P_{k}^{*}(x,s,
\tau ,x_{i})$ is a self--similar approximation corresponding to a quasifixed
point, and $\tau $ is an effective time necessary for reaching this
quasifixed point. The latter, in accordance with the inverse algebraic
transform (4), yields 
\begin{equation}
p_{k}^{*}(x,s,\tau ,x_{i})=x^{-s}P_{k}^{*}(x,s,\tau ,x_{i}).
\end{equation}

To illustrate these steps, consider an asymptotic expansion 
\begin{equation}
p_k(x,0) =\sum_{n=0}^k a_nx^n
\end{equation}
in the vicinity of $x_1=0$. Then, accomplishing the described procedure, for
Eq. (10) we find 
\begin{equation}
p_k^*(x,s,\tau,0) =\left [ p_{k-1}^{-k/s}(x,0) - 
\frac{ka_k\tau}{sa_0^{1+k/s}}x^k\right ]^{-s/k} .
\end{equation}
An important particular case is when $s\rightarrow\infty$, then Eq. (12)
gives 
\begin{equation}
\lim_{s\rightarrow\infty}p_k^*(x,s,\tau,0) = p_{k-1}(x,0)\exp\left ( \frac{
a_k}{a_0}\tau x^k\right ) .
\end{equation}
This shows how exponential functions naturally appear in our method,
together with the radical expressions of type (12).

An expression $p_k^*$, given either by (12) or by (13), as is seen, is a
function of a lower--order series $p_{k-1}$, 
\begin{equation}
p_k^* =F_k(p_{k-1}) .
\end{equation}
Analogously to the way by which we have come from an asymptotic series $p_k$
to the renormalized expression $p_k^*$, we can renormalize $p_{k-1}$
entering into relation (14), which gives 
\begin{equation}
p_k^{**} = F_k(p_{k-1}^*) = F_k(F_{k-1}(p_{k-2})) .
\end{equation}
Repeating such a renormalization $k$ times, we come to 
\begin{equation}
p_k^{*\ldots *} = F_k(F_{k-1}(\ldots F_1(p_0))\ldots ) .
\end{equation}

At each $n$--step of renormalization (14), two parameters, $s_n$ and $\tau_n$,
arise in the resulting expression, according to (12). Therefore, the 
$k$--times renormalized quantity (16) contains $2k$ such parameters, 
\begin{equation}
p_k^{*\ldots *} \equiv F_k^*(x,\bar s_k,\bar\tau_k, x_i) ,
\end{equation}
where the short--hand notation 
$$
\bar s_k\equiv \{ s_1,s_2,\ldots,s_k\} , \qquad \bar\tau_k\equiv
\{\tau_1,\tau_2,\ldots,\tau_k\} 
$$
is used.

The sets $\bar s_k$ and $\bar\tau_k$ are to be defined so that the
renormalization procedure would converge to a function satisfying the
boundary conditions (1) and (2). Suppose that we started from a series 
$p_k(x,x_i)$ written for an asymptotic region of $x_i$, with $i=1,2$.
Renormalizing this series $2k$ times, we get (17). In order that the
renormalized expression (17) could satisfy the correct asymptotic behaviour
at another boundary point $x_j$, with $j\neq i$, we have to require the
asymptotic condition 
\begin{equation}
F_k^*(x,\bar s_k,\bar\tau_k,x_i)\rightarrow p_k(x,x_j), \qquad x\rightarrow
x_j .
\end{equation}
Condition (18) defines the control sets 
\begin{equation}
\bar s_k =\bar s_k(x) , \qquad \bar\tau_k =\bar\tau_k(x)
\end{equation}
of control functions $s_1(x),s_2(x),\ldots,s_k(x)$, and $\tau_1(x),\tau_2(x),
\ldots,\tau_k(x)$. Substituting these control functions into (17), we 
obtain the final self--similar approximant 
\begin{equation}
f_k^*(x,x_i) = F_k^*(x,\bar s_k(x),\bar\tau_k(x),x_i) .
\end{equation}

Control functions are called so because of their role of controlling
convergence of the procedure to a function having the desired properties
[17]. In general, these functions are, really, functions of $x$, although in
particular cases they can become just parameters. In the latter case, they
can be called control parameters.

In order to make the above procedure transparent, let us consider a typical
case of two asymptotic expansions at $x_{1}=0$ and $x_{2}=\infty$. Assume
that at the left boundary we have a sequence 
\begin{equation}
p_{1}(x,0)=a_{0}+a_{1}x, \qquad p_{2}(x,0)=a_{0}+a_{1}x+a_{2}x^{2},\ldots
\end{equation}
of perturbative expansions $p_{k}(x,0)$, and at the right boundary, a
sequence 
\begin{equation}
p_{1}(x,\infty )=Ax^{n}, \qquad p_{2}(x,\infty )=Ax^{n}+Bx^{m},\ldots
\end{equation}
of asymptotic expressions $p_{k}(x,\infty )$, with $n\geq m$. Starting from 
$p_{1}(x,0)$, according to (12), we get 
\begin{equation}
p_{1}^{*}(x,s,\tau ,0)=\left( a_{0}^{-1/s}-\frac{a_{1}\tau }{sa_{0}^{1+1/s}}
x\right) ^{-s}.
\end{equation}
As the asymptotic boundary condition (18), we have 
\begin{equation}
p_{1}^{*}(x,s,\tau ,0)\rightarrow p_{1}(x,\infty ),\qquad x\rightarrow
\infty ,
\end{equation}
with $p_{1}(x,\infty )$ given by (22). Condition (24) holds true if and only
if 
\begin{equation}
s=-n,\qquad \tau =n\frac{a_{0}}{a_{1}}\left( \frac{A}{a_{0}}\right) ^{1/n}.
\end{equation}
Therefore, the first--order self--similar approximant, defined by (20),
becomes 
\begin{equation}
f_{1}^{*}(x,0)=\left( a_{0}^{1/n}+A^{1/n}x\right) ^{n}.
\end{equation}

Similarly, starting from $p_{2}(x,0)$ given in (21), we find the twice
renormalized expression 
\begin{equation}
F_{2}^{*}(x,s_{1},\tau _{1},s_{2},\tau _{2},0)=\left\{ \left[
p_{1}^{*}(x,s_{1},\tau _{1},0)\right] ^{-2/s_{2}}-
\frac{2a_{2}\tau _{2}}{s_{2}a_{0}^{1+2/s_{2}}}x^{2}\right\}^{-s_2/2} .
\end{equation}
Imposing the asymptotic boundary condition 
\begin{equation}
F_{2}^{*}(x,s_{1},\tau _{1},s_{2},\tau _{2},0)\rightarrow p_{1}(x,\infty
),\qquad x\rightarrow \infty ,
\end{equation}
we obtain 
\begin{equation}
s_{2}=-n,\qquad \tau _{2}=\frac{na_{0}}{2a_{2}}\left( \frac{A}{a_{0}}\right)
^{2/n}.
\end{equation}

Employing (29) for (27), we have 
\begin{equation}
F_{2}^{*}(x,s_{1},\tau _{1},s_{2},\tau _{2},0)=\left\{
[p_{1}^{*}(x,s_{1},\tau _{1},0)]^{2/n}+A^{2/n}x^{2}\right\}^{n/2} .
\end{equation}
The boundary condition 
\begin{equation}
F_{2}^{*}(x,s_{1},\tau _{1},s_{2},\tau _{2},0)\rightarrow p_{2}(x,\infty
),\qquad x\rightarrow \infty ,
\end{equation}
is satisfied provided that 
\begin{equation}
s_{1}=-\frac{n}{2}q,\qquad \tau _{1}=-s_{1}\frac{a_{0}}{a_{1}}\left( 
\frac{A}{a_{0}}\right) ^{2/nq}\left( \frac{2B}{nA}\right) ^{1/q},
\end{equation}
where the notation 
\begin{equation}
q\equiv 2+m-n<2\qquad (n>m)
\end{equation}
is used. With the control parameters given by (32), the function $p_{1}^{*}$
entering into (30) writes 
\begin{equation}
p_{1}^{*}(x,s_{1},\tau _{1},0)=\left[ a_{0}^{2/nq}+A^{2/nq}\left( 
\frac{2B}{nA}\right) ^{1/q}x\right] ^{nq/2}.
\end{equation}
Combining (30) with (34), we obtain the second--order self--similar
approximant 
\begin{equation}
f_{2}^{*}(x,0)=\left\{ \left[ a_{0}^{2/nq}+A^{2/nq}\left( \frac{2B}{nA}
\right) ^{1/q}x\right] ^{q}+A^{2/n}x^{2}\right\} ^{n/2},
\end{equation}
defined in Eq. (20). In the same way, we may proceed farther calculating a 
$k $--order self--similar approximant.

To complete this calculational procedure, we need to answer the following
question. Assume that we have two asymptotic expansions near two boundary
points. We may start from one of these expansions, say $p_{k}(x,x_{1})$,
imposing the boundary condition, as in (18), at another boundary point, in
this case at $x_{2}$, and thus obtaining the self--similar approximant 
$f_{k}^{*}(x,x_{1})$. The same procedure could be accomplished, starting from 
$p_{k}(x,x_{2})$ and imposing the boundary condition at $x_{1}$, thus
getting $f_{k}^{*}(x,x_{2})$. The question that arises is which of these two
approximants, $f_{k}^{*}(x,x_{1})$ or $f_{k}^{*}(x,x_{2})$ is expected to be
more accurate?

The answer to this question can be done from the point of view of stability
analysis [11-13]. To this end, let us take an expansion $p_{k}(x,x_{i})$
near a point $x_{i}$, with $i=1,2$. Suppose that $p_{0}(x,x_{i})$ depends on 
$x$. If $p_{0}(x,x_{i})$ does not depend on $x$, we have to take 
$p_{1}(x,x_{i})$. Define the expansion function $x(\varphi ,x_{i})$ by the
equation 
\begin{equation}
p_{0}(x,x_{i})=\varphi ,\qquad x=x(\varphi ,x_{i}).
\end{equation}
Introduce 
\begin{equation}
y_{k}(\varphi ,x_{i})=p_{k}(x(\varphi ,x_{i}),x_{i}),
\end{equation}
being a trajectory point of an approximation cascade $\{y_{k}^{i}\}$ formed
by the family of endomorphisms from (37). The stability of the cascade
trajectory is characterized by the local multipliers 
\begin{equation}
\mu _{k}(\varphi ,x_{i})\equiv 
\frac{\partial }{\partial \varphi }y(\varphi,x_{i}),
\end{equation}
whose images in the $x$--space are given by the local multipliers 
\begin{equation}
m_{k}(x,x_{i})\equiv \mu _{k}(p_{0}(x,x_{i}),x_{i})=\frac{\delta
p_{k}(x,x_{i})}{\delta p_{0}(x,x_{i})}=\frac{\partial
p_{k}(x,x_{i})/\partial x}{\partial p_{0}(x,x_{i})/\partial x}.
\end{equation}
The smaller absolute values $|m_{k}(x,x_{i})|$ of the multipliers correspond
to the more stable trajectory of the associated cascade, and the higher
stability implies the better convergence property of the related sequence of
approximations [12,13]. Therefore, in the asymptotic boundary condition
(18), we must choose that asymptotic expansion $p_{k}(x,x_{j})$ which
corresponds to the more stable cascade trajectory.

If two multipliers, $m_k(x,x_1)$ and $m_k(x,x_2)$, are equal or close to
each other, then we cannot decide apriori which of the self--similar
approximants, $f_k^*(x,x_1)$ or $f_k^*(x,x_2)$ is preferable. In such a
case, it is logical to define the average self--similar approximation 
\begin{equation}
f_k^*(x) \equiv\frac{1}{2}\left [ f_k^*(x,x_1) + f_k^*(x,x_2)\right ] .
\end{equation}
Usually, one of the approximations $f_k^*(x,x_i)$ where $i=1,2$, lies below,
and another above the exact function $f(x)$. In such a case, the errors of
these approximants compensate each other, essentially improving the accuracy
of the average approximant (40).

\section{Anharmonic Oscillators}

We start illustrating our interpolation approach with the models of
one--dimensional anharmonic oscillators described by the Hamiltonian 
\begin{equation}
H = -\frac{1}{2}\frac{d^2}{dx^2} + \frac{1}{2} x^2 + gx^m ,
\end{equation}
in which the space variable $x\in (-\infty,+\infty)$, the coupling parameter 
$g\in [0,\infty)$, and the power $m\geq 4$. These models are classical
touchstones from which everyone starts considering a new method.

Let us be interested in finding the ground--state energy $e(g)$ as a
function of the coupling parameter $g$. For this function, the asymptotic
expansions in the weak-- and strong--coupling limits are known. In the
weak--coupling limit, perturbation theory gives 
\begin{equation}
e_k(g,0) =\sum_{n=0}^k a_ng^n \qquad (g\rightarrow 0) .
\end{equation}
This series strongly diverges for any $g\neq 0$, since the coefficients $a_n$
grow like $n!$ as $n\rightarrow\infty$ [18,19]. The coefficients $a_n$ are,
of course, different for different types of oscillators, depending on $m$.
However, for the sake of simplicity, we do not use the double indexation. In
the strong coupling limit, one has [20] the expansion 
\begin{equation}
e_k(g,\infty) =\sum_{n=0}^k A_ng^{2(1-2n)/(m+2)} \qquad (g\rightarrow\infty).
\end{equation}
Here again the coefficients $A_n$ depend on $m$, that is, on the kind of
oscillator. Not marking this dependence explicitly will not lead to
confusion, since different kinds of oscillators will be considered
separately.

\vspace{3mm}

{\bf A. Quartic Oscillator}

\vspace{1mm}

Start with the quartic oscillator with $m=4$. For the first several
coefficients of the weak--coupling series (42), one has [18,19] 
$$
a_0=\frac{1}{2}, \qquad a_1 =\frac{3}{4}, \qquad a_2=-\frac{21}{8}, \qquad
a_3=\frac{333}{16} , \ldots 
$$
The coefficients of the strong--coupling expansion (43) have been computed
by many authors, starting from Hioe and Montroll [21]. One of the most
accurate computations have been accomplished by Weniger [22]. The values of 
the strong--coupling coefficients are $A_0=0.667986,\; A_1=0.143669,\;
A_2=0.008628,\; A_3=0.000818,\; A_4=0.000082,\; A_5=0.000008$.

Following the approach described in Sec.II, we may start from the
weak--coupling series (42) and define control functions from the asymptotic
condition (18) with $e_k(g,\infty)$ given by the strong--coupling expansion
(43). In the second order this gives 
$$
e_2^*(g,0) =\left [ a_0^6\left ( 1 +\frac{9a_1}{2a_0}\tau_0 g\right )^{4/3}
+ A^6g^2\right ]^{1/6} , 
$$
with the control parameter 
$$
\tau_0 =\frac{4A_0^4 A_1}{3a_0^4a_1}\left ( \frac{a_0^2}{6A_0A_1}
\right )^{1/4} = 0.66046 . 
$$
As a result, we obtain 
\begin{equation}
e_2^*(g,0) =\left [ a_0^6(1 +Cg)^{4/3} + A_0^6 g^2\right ] ^{1/6} ,
\end{equation}
where 
\[
C =\frac{6A_0^4A_1}{a_0^5}\left (\frac{a_0^2}{6A_0A_1}\right )^{1/4}. 
\]

In the same way, starting from the strong--coupling expansion (43) and
defining control functions from asymptotic condition (18) at the left
boundary, we find 
\begin{equation}
e_2^*(g,\infty) =\left [ a_0^4 + \left ( 16a_0^6a_1^2 + A_0^8g^{2/3}
\right )^{1/2}g\right ]^{1/4} .
\end{equation}
Numerical calculations show that both approximants (44) and (45) are close
to each other. The accuracy of a self--similar approximant $e_k^*(g,x_i)$
can be estimated by comparing the values, it gives for different coupling
parameters $g$, with precise numerical calculations accomplished by Hioe
and Montroll [21] for $g$ in the interval $0.02\leq g\leq 20000$. The
maximal percentage error of the left approximant (44) is $-2.9\%$
occurring at $g=0.3$, and the largest error of (45) is $4.2\%$ at $g=2$.
For all $g$, the left approximant (44) lies below the exact values of the
energy, while the right approximant (45) is above the exact values. The
average self--similar approximant 
$$
e_2^*(g) =\frac{1}{2}\left [ e_2^*(g,0) + e_2^*(g,\infty)\right ] 
$$
has the maximal error of $1.4\%$ at $g=2$.

As we see, quite simple analytical expression provide sufficiently good
accuracy, with the maximal error around one percent. As far as the structure
of perturbative series for the quartic anharmonic oscillator is analogous to
that of series for the so--called $\varphi^4$ model of quantum field theory
[23], we may hope that for the latter one also could construct analogous
self--similar approximants.

\vspace{3mm}

{\bf B. Sextic Oscillator}

\vspace{1mm}

The sextic oscillator $(m=6)$ is interesting being a borderline case between
the models whose perturbative series are Pad\'e summable and those whose
series cannot be summed. For the sextic oscillator, Pad\'e approximants
converge so slowly that they are computationally useless [5,24].

Employing the approach of Sec.II, we use the coefficients $A_0=0.680703,\; 
A_1=0.129464,\; A_2=0.005512.\; A_3=0.000328,\; A_4=0.000018,\; A_5=0.000001$.
We find the first self--similar approximants, from the left, 
\begin{equation}
e_1^*(g,0) =\frac{1}{2}\left ( 1 + 16A_0^4 g\right )^{1/4} ,
\end{equation}
and from the right, 
\begin{equation}
e_1^*(g,\infty) =\frac{1}{2}\left ( 1+ 4A_0^2g^{1/2}\right )^{1/2} .
\end{equation}
The maximal error of Eq. (46), with respect to accurate numerical results
[5] that can be treated as exact, is about $-11\%$, and the maximal error of
(47) is about $8\%$.

In the second order, we find 
\begin{equation}
e_2^*(g,0) =\frac{1}{2}\left [ ( 1+ 2Cg)^{3/2} + (2A_0)^8g^2\right ]^{1/8} ,
\end{equation}
where $C=3.428$, with the maximal error $-6\%$. The right approximant has a
comparable accuracy.

The third--order left approximant is 
\begin{equation}
e_3^*(g,0) =\frac{1}{2}\left\{\left [ (1+2B_1g)^{3/2} + 4B_2 g^2
\right ]^{5/4} + (2A_0)^{12} g^3\right\}^{1/12} ,
\end{equation}
where $B_1=4.831$ and $B_2=9.352$. The maximal error is around $-4\%$.

The fourth--order approximant can be written as 
\begin{equation}
e_4^*(g,0) =\frac{1}{2}\left\{\left [\left [ (1+2C_1g)^{3/2} + 4C_2g^2
\right ]^{5/4} + 8C_3g^3\right ]^{7/6} +(2A_0)^{16}g^4\right\}^{1/16} ,
\end{equation}
with $C_1=6.078,\; C_2=18.143$, and $C_3=22.322$. The maximal error is about 
$-3\%$.

In the fifth order, we find 
\begin{equation}
e_5^*(g,0) =\frac{1}{2}\left\{\left [\left [\left [ (1+2D_1g)^{3/2} +
4D_2g^2\right ]^{5/4} + 8D_3g^3\right ]^{7/6} + 16D_4g^4\right ]^{9/8} +
(2A_0)^{20}g^5\right\}^{1/20} ,
\end{equation}
where $D_1=7.215,\; D_2=28.848,\; D_3=56.001$, and $D_4=49.39$. The maximal
error is $-2.5\%$.

For the sixth order, we obtain 
$$
e_6^*(g,0) = \frac{1}{2}\left\{\left [\left [\left [\left [ (1+ 2K_1g)^{3/2}
+ 4K_2g^2\right ]^{5/4} + 8K_3g^3\right ]^{7/6} + 16K_4g^4\right ]^{9/8} +
32K_5g^5\right ]^{11/10} + \right. 
$$
\begin{equation}
\biggl.+\bigl ( 2A_0 \bigr )^{24}g^6\biggr\}^{1/24} ,
\end{equation}
with the coefficients $K_1=8.256,\; K_2=41.122,\; K_3=109.122,\; 
K_4=153.119$, and $K_5=104.156$. The maximal error of approximant (52) is 
$-2\%$.

Eqs. (46)-(52) show that the accuracy of the self--similar approximants
improves with increasing order. To demonstrate that there is uniform
numerical convergence for all $g$, we present in Table I the percentage
errors $\varepsilon_k^*(g,x_i)$ of the corresponding approximants 
$e_k^*(g,x_i)$, as compared to exact values $e(g)$. The accuracy in each
order can also be improved by defining the average approximants (40). We
show this for the case of the approximant 
\begin{equation}
e_1^*(g) =\frac{1}{2}\left [ e_1^*(g,0) + e_1^*(g,\infty)\right ] ,
\end{equation}
whose errors are also presented in Table I.

\vspace{3mm}

{\bf C. Octic Oscillator}

\vspace{1mm}

The case of the octic oscillator $(m=8)$ is important to consider
remembering that Pad\'e approximants are not able to sum the corresponding
perturbation series [24,25]. As we show below, in our approach we obtain a
series of self--similar approximants exhibiting uniform numerical
convergence.

Here we use the following coefficients $A_0=0.704046,\; A_1=0.120626,\;
A_2=0.004168,\; A_3=0.000188,\; A_4=0.000007,\; A_5=0.000001$. The
first--order left approximant reads 
\begin{equation}
e_1^*(g,0) =\frac{1}{2}\left ( 1 + 32 A_0^5g\right )^{1/5} ,
\end{equation}
while the right approximant is 
\begin{equation}
e_1^*(g,\infty) =\frac{1}{2}\left ( 1 +4A_0^2g^{2/5}\right )^{1/2} .
\end{equation}
Comparing this with numerical results [5], we find that the maximal error of
(54) is about $-13\%$ and that of (55) is $8\%$.

For the second--order left approximant we have 
\begin{equation}
e_2^*(g,0) =\frac{1}{2}\left [ (1 +2Cg)^{8/5}+(2A_0)^{10} g^2\right ]^{1/10},
\end{equation}
with $C=5.944$, the maximal error being $-8\%$.

In the third order, we get 
\begin{equation}
e_3^*(g,0) =\frac{1}{2}\left\{\left [ ( 1+2B_1g)^{8/5} + 4B_2g^2\right %
]^{13/10} + (2A_0)^{15} g^3\right \}^{1/5} ,
\end{equation}
where $B_1=8.671$ and $B_2=26.807$. The maximal error is $-6\%$.

The fourth--order approximant is 
\begin{equation}
e_4^*(g,0) =\frac{1}{2}\left\{\left [\left [ (1+2C_1g)^{8/5} + 4C_2g^2\right %
]^{13/10} + 8C_3g^3\right ]^{6/5} + (2A_0)^{20}g^4\right\}^{1/20} ,
\end{equation}
with $C_1=11.151,\; C_2=55.077$, and $C_3=104.667$. The maximal error is 
$-4.5\%$.

The fifth--order approximant writes 
\begin{equation}
e_5^*(g,0) =\frac{1}{2}\left\{\left [\left [\left [ (1+2D_1g)^{8/5} +
4D_2g^2\right ]^{13/10} + 8D_3g^3\right ]^{6/5} + 16D_4g^4\right ]^{23/20} +
(2A_0)^{25} g^5\right\}^{1/25} ,
\end{equation}
where $D_1=13.443,\; D_2=91.126,\; D_3=282.775$, and $D_4=377.013$. The
maximal error is $-3.5\%$.

For the sixth order, we obtain 
$$
e_6^*(g,0) =\frac{1}{2}\left\{\left [\left [\left [\left [ (1 +2K_1g)^{8/5}
+ 4K_2g^2\right ]^{13/10} + 8K_3g^3\right ]^{6/5} + 16K_4g^4\right ]^{23/20}
+ 32K_5g^5\right ]^{28/25} +\right. 
$$
\begin{equation}
\biggl. + (2A_0)^{30} g^6\biggr\}^{1/30} ,
\end{equation}
where $K_1=15.508,\; K_2=133.486,\; K_3=581.021,\; K_4=1274$, and 
$K_5=1291$. The maximal error is $-3\%$.

As we see, in our approach there is no principal difference between the
types of oscillators, whether it is quartic, sextic or octic; for each of
them we can easily construct a uniformly convergent sequence of
self--similar approximants. The accuracy of the latter in each order can be
essentially improved by composing average approximants, as in (53). The
errors of the obtained approximants are collected in Table II.

Let us emphasize that our aim here was to derive {\it analytical} formulas.
The approximants we have constructed are easier to use than more complicated
expressions that follow from renormalized perturbation theory [17], in which
control functions are introduced into a zero--order Hamiltonian [6,7,26-31].
This especially concerns the sextic and octic oscillators. We think, that
the possibility to have simple and accurate formulas, valid for the whole
range of coupling parameters is an advantage of our approach that could be a
useful tool for analysing the properties of quantum dots [32-34].

\section{Double--Well Oscillators}

Models with double--well potentials are notorously known to be difficult for
approximate treatment. For instance, perturbation theory in this case
results in series that are not Pad\'e summable. At the same time such
potentials are quite common for various problems encountered in physics and
chemistry (see discussion in Ref.[35]).

One of the difficulties of dealing with double--well models is that the
corresponding physical quantities, as functions of the coupling parameter,
can display not two characteristic regions of behaviour, that is, the
weak--coupling and the strong--coupling regions, but a third region,
intermediate between weak and strong coupling. This behaviour is similar to
that of some models of quantum field theory where in the transition region
instanton effects are crucial, bridging the weak and strong coupling limits
[36,37].

\vspace{3mm}

{\bf A. Zero--Dimensional Model}

\vspace{1mm}

Let us, first, consider the so--called zero--dimensional double--well model
whose free energy is written as 
\begin{equation}
f(g) =-\ln Z(g) ,
\end{equation}
where 
\begin{equation}
Z(g) =\frac{1}{\sqrt{\pi}}\int_{-\infty}^{+\infty} \exp\left\{ -H(x)
\right\} dx
\end{equation}
plays the role of a partition function with the Hamiltonian 
\begin{equation}
H(x) = -x^2 + gx^4 , \qquad g\geq 0 .
\end{equation}
The latter has a maximum $H(0)=0$ at $x=0$ and two minima $H(\pm a) = -1/4g$
at $x=\pm a=\pm 1/\sqrt{2g}$.

Direct use of perturbation theory, in powers of $g$, to the free energy (61)
is impossible, since $f(g)\rightarrow -\infty$ as $g\rightarrow 0$. Thence,
a special procedure is necessary. To this end, we define the trial
Hamiltonians

\begin{equation}
H_\pm (x) =\omega^2(x\pm a)^2 - u_0 ,
\end{equation}
in which 
$$
a=\frac{1}{\sqrt{2g}} , \qquad u_0 =\frac{1}{4g} , 
$$
and $\omega$ is a trial parameter to become latter a control function. The
partition function (62) can be written in the form 
\begin{equation}
Z(g) =\frac{1}{2\sqrt{\pi}}\int_{-\infty}^{+\infty} \exp\left\{ - H_+(x) -
\Delta H_+(x)\right\} dx + 
\frac{1}{2\sqrt{\pi}}\int_{-\infty}^{+\infty}\exp\left\{ - H_-(x) - \Delta
H_-(x)\right\} dx ,
\end{equation}
where 
$$
\Delta H_\pm = H(x) - H_\pm (x) . 
$$

The free energy (61) is expanded in powers of $\Delta H_\pm(x)$, with the
zero--order term 
\begin{equation}
F_0(g,\omega) =\ln\omega -u_0 ,
\end{equation}
the first--order term 
\begin{equation}
F_1(g,\omega) =\ln\omega - a^2 -\frac{1}{2\omega^2} -\frac{1}{2} + \left (
a^4+\frac{3a^2}{\omega^2} +\frac{3}{4\omega^4}\right ) g ,
\end{equation}
and so on.

The control function $\omega(g)$ is defined from the quasifixed--point
condition 
\begin{equation}
\frac{\partial}{\partial\omega} F_1(g,\omega) = 0 ,
\end{equation}
which gives 
\begin{equation}
\omega(g) =\left (\frac{3g}{\sqrt{1+3g}-1}\right )^{1/2} .
\end{equation}
This control function in the weak--coupling limit, as $g\rightarrow 0$,
behaves as 
\begin{equation}
\omega(g)\simeq\sqrt{2}\left ( 1 +\frac{3}{8} g -\frac{45}{128} g^2\right ) ,
\end{equation}
and in the strong--coupling limit, as $g\rightarrow\infty$, it has the
asymptotic behaviour 
\begin{equation}
\omega(g)\simeq (3g)^{1/4} +\frac{1}{2}(3g)^{-1/4} +\frac{1}{8}(3g)^{-3/4} .
\end{equation}
For $g\in [0,\infty)$, function (69) changes in the interval $\sqrt{2}%
\leq\omega(g) <\infty$.

Defining 
\begin{equation}
f_k(g)\equiv F_k(g,\omega(g)) ,
\end{equation}
from (66) and (67) we have 
\begin{equation}
f_0(g) =\ln\omega -\frac{1}{4g} , \qquad
f_1(g) =\ln\omega -\frac{1}{4g} -\frac{1}{4} +\frac{1}{2\omega^2} .
\end{equation}
Similarly, calculating $F_2(g,\omega)$, we come in the second order to 
\begin{equation}
f_2(g) =\ln\omega -\frac{1}{4g} -\frac{1}{3} -\frac{5}{3\omega^2} + 
\frac{14}{3\omega^4} .
\end{equation}

Wishing to estimate the accuracy of the approximations $f_1(g)$ and 
$f_2(g)$, let us notice that the exact function (61) changes from $-\infty$ 
as $g\rightarrow 0$ to $+\infty$ as $g\rightarrow\infty$, crossing zero at 
$g=g_c = 2.758$, that is, $f(g_c)=0$. Therefore, we cannot define the
percentage error in the standard way as $100\%\times [f_k(g)-f(g)]/f(g)$,
since such a definition contains zero in the denominator. Instead of this,
we may evaluate the accuracy of a crossing point given by the corresponding
approximation, that is, the accuracy of the solution $g_c^{(k)}$ to the
equation 
\begin{equation}
f_k(g_c^{(k)}) = 0 .
\end{equation}
For the first approximation, we have $g_c^{(1)}= 0.585$ which gives the
error of $-79\%$, as compared to the exact $g_c=2.758$. For the second
approximation, we get $g_c^{(2)}= 1.352$, whose error is $-51\%$. As is
seen, this accuracy is not high, so that it is desirable to improve it.

Introduce the function 
\begin{equation}
\alpha(g)\equiv\frac{1}{\omega^2(g)} =\frac{\sqrt{1+3g}-1}{3g} .
\end{equation}
This function changes in the interval $0\leq\alpha(g)\leq\frac{1}{2}$, with
the asymptotic behaviour 
$$
\alpha(g)\simeq \frac{1}{2} -\frac{3}{8}g \qquad (g\rightarrow 0) , 
$$
$$
\alpha(g)\simeq \frac{1}{\sqrt{3g}}-\frac{1}{3g} \qquad 
(g\rightarrow\infty) . 
$$
Using (76), we can write for the approximations $f_k(g)$ the following
expressions: in the zero order, 
\begin{equation}
f_0(g) =-\frac{1}{2}\ln\alpha -\frac{3\alpha^2}{4(1-2\alpha)} ,
\end{equation}
where $\alpha=\alpha(g)$, in the first order, 
\begin{equation}
f_1(g) = f_0(g) -\frac{1}{4} +\frac{1}{2}\alpha ,
\end{equation}
and in the second order, 
\begin{equation}
f_2(g) = f_0(g) -\frac{1}{3}\left ( 1 +5\alpha -14\alpha^2\right ) .
\end{equation}

The zero--order approximation (77) correctly describes the weak and strong
coupling limits of the exact function (61), but it is not accurate in the
intermediate region. In this region, the behaviour of the approximation 
$$
f_k(g) = f_0(g) + p_k(\alpha) 
$$
is governed by a series $p_k(\alpha)$ in powers of $\alpha=\alpha(g)$.
Renormalizing this series twice, according to the bootstrap procedure [16],
we obtain 
\begin{equation}
f_2^*(g) = f_0(g) -\frac{1}{3}\exp\left\{ 5\alpha\exp\left ( - 
\frac{14}{5} \alpha\right )\right\} ,
\end{equation}
where $\alpha=\alpha(g)$ is given by (76). The obtained self--similar
approximant (80) provides much better approximation to function (61),
compared to $f_1(g)$ and $f_2(g)$. The crossing point $g_c^* = 2.858$,
defined by the condition $f_2^*(g_c^*) = 0$, is quite close to the exact 
$g_c$ and gives an error $3.6\%$. That Eq.(80) is an accurate approximant 
is also clearly seen in Fig. 1.

\vspace{3mm}

{\bf B. One--Dimensional Oscillator}

From the model of the previous section, we now pass to a more realistic case
of the double--well oscillator with the Hamiltonian 
\begin{equation}
H = -\frac{1}{2}\frac{d^2}{dx^2} +\frac{1}{16g} -\frac{1}{2}x^2 +gx^4 ,
\end{equation}
in which $x\in(-\infty,+\infty)$ and $g\in[0,\infty)$. The problem of
finding the eigenvalues of Hamiltonian (81) is a challenge for any
analytical method, although there are several numerical techniques
calculating the eigenvalues with reasonable accuracy [38-43]. It is
especially difficult to calculate the lowest energy levels. The main problem
here is that instanton contributions are crucial in the weak--coupling
region providing for an exponentially small splitting of energy levels. In
addition, the energy of the ground--state level is not a monotonous function
of the coupling parameter $g$, which is also related to the instanton
contributions. Below we shall consider the most difficult case of the
ground--state energy and that of the first excited level separated from the
former, in the weak--coupling region, by an exponentially small gap.

To construct interpolation formulas, we need asymptotic expansions for the
weak and strong coupling limits. We shall use such expansions derived in
Ref.[44]. The ground--state energy $e_+(g)$ corresponds to a symmetric wave
function, while the first excited level, with an energy $e_-(g)$,
corresponds to an antisymmetric wave function. These energies can be written
in the form 
\begin{equation}
e_\pm(g) =\bar e(g) \mp\frac{1}{2}\Delta(g) ,
\end{equation}
in which 
\begin{equation}
\bar e(g) \equiv\frac{1}{2}\left [ e_+(g) + e_-(g)\right ] , \qquad
\Delta(g) \equiv e_-(g) -e_+(g) .
\end{equation}
Let us notice that the Hamiltonian (81) is shifted, as compared to the
standard form, by the term $1/16g$, which makes the spectrum of (81)
everywhere positive [44].

For the average energy and the gap, defined in (83), we have [44] in the
weak--coupling limit, when $g\rightarrow 0$, 
\begin{equation}
\bar e(g)\simeq \frac{1}{\sqrt{2}} -\frac{21}{64}g
\end{equation}
and, respectively, 
\begin{equation}
\Delta(g)\simeq \frac{a}{g}\exp\left (\frac{b}{g} + c\right ) ,
\end{equation}
where 
$$
a=\frac{303}{1024}, \qquad b=-\frac{\sqrt{2}}{4}, \qquad c=\frac{9}{4} . 
$$

In the strong--coupling limit, when $g\rightarrow\infty$, we may derive [44]
for the energies 
$$
e_+(g)\simeq \frac{3}{8}(6g)^{1/3} -\frac{1}{4}(6g)^{-1/3} + 
\frac{13}{12} (6g)^{-1} -\frac{2705}{3456}(6g)^{-5/3} , 
$$
$$
e_-(g)\simeq \frac{9}{8}(10)^{1/3} -\frac{3}{4}(10g)^{-1/3} + 
\frac{377}{144} (10g)^{-1} -\frac{159139}{31104}(10g)^{-5/3} . 
$$
From here, for the average energy we get 
\begin{equation}
\bar e(g) \simeq Ag^{1/3} + Bg^{-1/3} + Cg^{-1} + Dg^{-5/3} ,
\end{equation}
where 
$$
A=1.552580, \qquad B=-0.242850, \qquad C=0.221181, \qquad D=-0.074868 . 
$$
And for the gap, we find 
\begin{equation}
\Delta(g)\simeq A_1g^{1/3} + B_1g^{-1/3} + C_1g^{-1} +D_1g^{-5/3} ,
\end{equation}
where 
$$
A_1=1.742319, \qquad B_1 =-0.210539, \qquad C_1=0.081250, \qquad
D_1=-0.070721. 
$$

Constructing a self--similar approximation from the right to the left, we
have the right approximant 
\begin{equation}
e_4^*(g,\infty) = \bar e^*(g) \mp\frac{1}{2}\Delta^*(g) ,
\end{equation}
in which 
\begin{equation}
\bar e^*(g) =\left [\frac{1}{4} + A^4g^{4/3}\exp\left ( \frac{4B}{Ag^{2/3}}
\right )\right ]^{1/4} + \frac{Dg}{\left [ g^{2/3} +\left (\frac{64}{21}
|D|\right )^{1/4}\right ]^4} .
\end{equation}
The most difficult here is to interpolate between the power--law expansion
(87) for the gap, in the strong coupling limit, and the exponential
behaviour (85) in the weak--coupling region. Nevertheless, employing the
technique of Sec. II, we obtain for the gap the form 
\begin{equation}
\Delta^*(g) =\frac{\alpha^*(g)}{g}\exp\left\{\frac{\beta^*(g)}{g} \right\} ,
\end{equation}
describing a renormalized instanton contribution, where 
$$
\alpha^*(g) =\left [ a^{3/2} + A_1^{3/2}g^2\exp\left\{ 
\frac{3B_1}{2A_1g^{3/2}}\exp\left (\frac{C_1}{B_1g^{2/3}}\right ) 
\right\}\right ]^{2/3},
$$
$$
\beta^*(g) =\frac{|b|D_1}{\left [ |D_1|^{4/3} +(|b|A_1)^{4/3} (\tau
+g^{2/3})^{1/2}g\right ]^{3/4}} , \qquad
\tau\equiv 9\left (\frac{D_1^4}{A_1^4|b|^7}\right )^{2/3} = 0.22417 . 
$$

The behaviour of two branches of Eq. (88), compared with the exact numerical 
data, correctly describes the nonmonotonic behaviour of the ground--state
energy and the exponential branching at $g=0$. The accuracy of the 
constructed formulas is very good for both the instanton--dominated region 
($g\ll 0.2$) and instanton--free region ($g\gg 0.2$), with the error
tending to zero in both these limits. The most difficult for description
is a narrow intermediate region around the point $g\approx 0.2$, where an
error is about $25\%$. The accuracy can be improved taking into account
more terms of asymptotic expansions, but then the formulas become
essentially more complicated.

\vspace{3mm}

{\bf C. Quasistationary States}

\vspace{1mm}

Quasistationary or resonance states are encountered in a variety of studies
in atomic and molecular physics. A good discussion and many references are
given in Ref. [45]. Several numerical calculations are known for this
problem [6,42,45,46]. Here we derive analytical formulas for both the real
and imaginary parts of the spectrum of the Hamiltonian 
\begin{equation}
H = -\frac{1}{2}\frac{d^2}{dx^2} +\frac{1}{2} x^2 -gx^4 ,
\end{equation}
in which $x\in(-\infty,+\infty)$ and $g\in[0,\infty)$.

First, we have to analyse the asymptotic behaviour of the spectrum in the
weak and strong coupling limits. To this end, we invoke perturbation theory
starting from the Hamiltonian 
\begin{equation}
H_0=-\frac{1}{2}\frac{d^2}{dx^2} +\frac{u^2}{2} x^2 ,
\end{equation}
in which $u$ is a trial parameter to be converted into a control function.
Introducing, for convenience, the notation 
\begin{equation}
E_k(g,u) \equiv\left ( n +\frac{1}{2}\right ) F_k(g,u)
\end{equation}
of a $k$--order perturbative expression for the energy of a level $%
n=0,1,2,\ldots$, we find 
$$
F_0(g,u) = u , \qquad
F_1(g,u)= u-\frac{u}{4}(2\alpha -\beta) , 
$$
\begin{equation}
F_2(g,u) = F_1(g,u) -\frac{u}{8}(\alpha^2 -2\alpha\beta + 2\beta^2\delta) ,
\end{equation}
and so on, analogously to the case of the anharmonic oscillator [13,31],
with the notation 
$$
\alpha\equiv 1-\frac{1}{u^2} , \qquad \beta \equiv -\frac{6\gamma g}{u^3},
\qquad \gamma\equiv\frac{n^2+n+1/2}{n+1/2} , \qquad \delta\equiv \frac{%
17n^2+17n+21}{(6\gamma)^2} . 
$$

The control function $u=u(g)$ can be defined from the fixed--point condition 
\begin{equation}
\frac{\partial}{\partial u} F_k(g,u) = 0.
\end{equation}
In the first order, we get the equation 
\begin{equation}
u^3 - u +6\gamma g = 0 ,
\end{equation}
as a result of which 
$$
\alpha=\beta=\frac{u^2-1}{u^2} . 
$$

Substituting the solution $u(g)$ to Eq. (96) into (94), we define 
\begin{equation}
f_k(g)\equiv F_k(g,u(g)) .
\end{equation}
Then, from (94) we have 
\begin{equation}
f_1(g) =\frac{3}{4}u +\frac{1}{4u} , \qquad
f_2(g) = f_1(g) +\frac{1}{8}( 1- 2\delta)\alpha^2 u .
\end{equation}

From three solutions of Eq. (96) we need to choose that which satisfies the
boundary condition $u(g)\rightarrow 1$, as $g\rightarrow 0$. Such a solution
is 
\begin{equation}
u(g) =\left (\frac{1}{9}\sqrt{729\gamma^2g^2 -3} -3\gamma g\right )^{1/3} + 
\frac{1}{3}\left (\frac{1}{9}\sqrt{729\gamma^2g^2-3} -3\gamma g \right
)^{-1/3} .
\end{equation}
The control function (99) is real for $g\leq g_n$, where 
\begin{equation}
g_n\equiv\frac{g_0}{\gamma} =\frac{n+1/2}{n^2+n+1/2}g_0 , \qquad g_0\equiv%
\frac{\sqrt{3}}{27} = 0.064150 .
\end{equation}
Complex roots of (99) appear only after $g>g_n$. The fact that ${\rm Im}\;
u_k(g)=0$ for $g\leq g_n$ leads to ${\rm Im}\; f_k(g)=0$, when $g\leq g_n$.

Asymptotic expansions in the weak and strong coupling limits can be written
for arbitrary energy levels. For illustrative purpose, we shall write down
expansions for the ground state $(n=0)$ and the first excited state $(n=1)$.

For the ground state, when $n=0$ and $\gamma=1$, function (98) yields 
for the real parts 
$$
{\rm Re}\; f_1(g)\simeq 1 -\frac{3}{2} -\frac{9}{2}g^2 -27g^3 - \frac{1701}{8%
} g^4 , 
$$
\begin{equation}
{\rm Re}\; f_2(g)\simeq 1 -\frac{3}{2}g -\frac{21}{4} g^2-\frac{153}{4}g^3 - 
\frac{729}{2} g^4 \qquad (n=0) ,
\end{equation}
while the imaginary parts are 
\begin{equation}
{\rm Im}\; f_1(g) ={\rm Im}\; f_2(g) = 0 , \qquad g\rightarrow 0 .
\end{equation}

For the first excited level, for which $n=1$ and $\gamma=5/3$, 
for the real parts of (98) we find 
$$
{\rm Re}\; f_1(g)\simeq 1 -\frac{5}{2}g -\frac{25}{2}g^2 -125g^3 - 
\frac{13125}{8} g^4 , 
$$
\begin{equation}
{\rm Re}\; f_2(g)\simeq 1 -\frac{5}{2}g -\frac{55}{4} g^2-\frac{625}{4}g^3 - 
\frac{9375}{4} g^4 \qquad (n=1) ,
\end{equation}
while their imaginary parts are again as in (102).

In the strong--coupling limit, when $g\rightarrow\infty$, for the ground 
state we obtain the real parts 
$$
{\rm Re}\; f_1(g)\simeq \frac{3}{8}(6g)^{1/3} +\frac{1}{4}(6g)^{-1/3} + 
\frac{1}{12}(6g)^{-1} +\frac{1}{108}(6g)^{-5/3}-\frac{1}{648}(6g)^{-7/3}, 
$$
\begin{equation}
{\rm Re}\; f_2(g)\simeq \frac{35}{96}(6g)^{1/3} +\frac{77}{288}(6g)^{-1/3} + 
\frac{17}{144}(6g)^{-1} +\frac{43}{1944}(6g)^{-5/3} - 
\frac{211}{23328} (6g)^{-7/3} ,
\end{equation}
and the imaginary parts 
$$
{\rm Im}\; f_1(g)\simeq -\frac{3\sqrt{3}}{8}(6g)^{1/3} + 
\frac{\sqrt{3}}{4} (6g)^{-1/3} - \frac{\sqrt{3}}{108}(6g)^{-5/3}
-\frac{\sqrt{3}}{648} (6g)^{-7/3} , 
$$
\begin{equation}
{\rm Im}\; f_2(g)\simeq - \frac{35\sqrt{3}}{96}(6g)^{1/3} + 
\frac{77\sqrt{3}}{288}(6g)^{-1/3} -\frac{43\sqrt{3}}{1944}(6g)^{-5/3} - 
\frac{211\sqrt{3}}{23328}(6g)^{-7/3} .
\end{equation}

For the strong--coupling limit, in the case of the first excited level 
$(n=1)$, we find the real parts 
$$
{\rm Re}\; f_1(g)\simeq \frac{3}{8}(10g)^{1/3} +\frac{1}{4}(10g)^{-1/3} +
\frac{1}{12}(10g)^{-1} +\frac{1}{108}(10g)^{-5/3} - 
\frac{1}{648} (10g)^{-7/3} , 
$$
\begin{equation}
{\rm Re}\; f_2(g)\simeq \frac{59}{160}(10g)^{1/3} +\frac{25}{96}(10g)^{-1/3}
+\frac{5}{48}(10g)^{-1} +\frac{11}{648}(10g)^{-5/3} - \frac{47}{7776}%
(10g)^{-7/3} ,
\end{equation}
and imaginary parts 
$$
{\rm Im}\; f_1(g)\simeq -\frac{3\sqrt{3}}{8}(10g)^{1/3} +
\frac{\sqrt{3}}{4} (10g)^{-1/3} -\frac{\sqrt{3}}{108}(10g)^{-5/3} -
\frac{\sqrt{3}}{648} (10g)^{-7/3} , 
$$
\begin{equation}
{\rm Im}\; f_2(g)\simeq -\frac{59\sqrt{3}}{160}(10g)^{1/3} +
\frac{25\sqrt{3}}{96}(10g)^{-1/3} -\frac{11\sqrt{3}}{648}(10g)^{-5/3} - 
\frac{47\sqrt{3}}{7776}(10g)^{-7/3} .
\end{equation}

The accuracy of the real parts of the approximations in (98) is sufficiently
good; the maximal error in the first order is $-3.3\%$ and that of the
second order is $2\%$. In the case of the corresponding imaginary parts, the
maximal error for $g>g_n$ is on the order of $10\%$. However, for $g\leq g_n$%
, imaginary parts are identically zero, because of which their
weak--coupling expansions do not exist.

To correct the described deficiency, let us consider the fixed--point
condition (95) of second order. Then, the equation for the control function
becomes of the form 
\begin{equation}
u^6 -2u^4+16\gamma gu^3 + u^2 -16\gamma gu + 120\gamma^2\delta g^2 = 0 .
\end{equation}
For the ground--state level, for which $n=0,\;\gamma=1$, and $\delta=7/12$,
Eq. (108) reduces to 
\begin{equation}
u^6 -2u^4 +16gu^3 +u^2 -16gu+70g^2 = 0 .
\end{equation}
The solution to (109) will be denoted by $u^*(g)$, in order to distinguish
it from that of (96). Among the solutions of Eq. (109) it is necessary to
choose that which satisfies the asymptotic boundary condition $%
u^*(g)\rightarrow 1$, as $g\rightarrow 0$. Substituting $u^*(g)$ into 
$$
F_2(g,u) =\frac{3}{8}u +\frac{3}{4u} -\frac{3g}{u^2} -\frac{1}{8u^3} + \frac{%
3g}{2u^4} -\frac{21g^2}{4u^5} , 
$$
we come to 
\begin{equation}
f_2^*(g)\equiv F_2(g,u^*(g)) .
\end{equation}
Defining the real and imaginary parts of (110), we take those of them for
which 
$$
{\rm Re}\; f_k^*(g) > 0 , \qquad {\rm Im}\; f_k^*(g) \leq 0 . 
$$

To solve Eq.(109), we introduce a new variable 
\begin{equation}
\lambda\equiv u(u^2-1) ,
\end{equation}
for which (109) reduces to a much simpler equation 
\begin{equation}
\lambda^2 +16g\lambda + 70g^2 = 0 .
\end{equation}
From two solutions to this equation, $\lambda_{1,2} = - (8\pm i\sqrt{6}) g$, 
we need to choose such that, together with Eq. (111), gives the control
function $u^*(g)$ satisfying the conditions discussed above. The desired
solution to (112) is 
\begin{equation}
\lambda(g) =-(8+i\sqrt{6})g .
\end{equation}
Then from the equation $u^3-u-\lambda=0$ we find

\begin{equation}
u^*(g)= \left (\frac{\lambda}{2} +\frac{1}{18}\sqrt{81\lambda^2-12}
\right )^{1/3} + \frac{1}{3} \left (\frac{\lambda}{2} +\frac{1}{18}
\sqrt{81\lambda^2-12}\right )^{-1/3} ,
\end{equation}
where $\lambda=\lambda(g)$ is defined in (113).

In the weak--coupling limit, when $g\rightarrow 0$ together with 
$\lambda\rightarrow 0$, approximant (110) has the following asymptotic 
expansions for the real part 
\begin{equation}
{\rm Re}\; f_2^*(g)\simeq 1-\frac{3}{2}g -\frac{21}{4}g^2 -41g^3 - 
\frac{14157}{32}g^4 -5547g^5 -\frac{9441289}{128}g^6 ,
\end{equation}
and for the imaginary part 
\begin{equation}
{\rm Im}\; f_2^*(g)\simeq -\frac{3}{4}\sqrt{6}g^3 -27\sqrt{6}g^4 -
\frac{1413}{2}\sqrt{6}g^5 -\frac{64779}{4}\sqrt{6}g^6 .
\end{equation}

In the strong--coupling limit, when $g\rightarrow\infty$ and 
$|\lambda|\rightarrow\infty$, for the real part of (110), we have 
$$
{\rm Re}\; f_2^*(g)\simeq 0.672436g^{1/3} + 0.145202g^{-1/3} +
0.016735g^{-1}+0.000603g^{-5/3}- 
$$
\begin{equation}
-0.000088g^{-7/3} - 0.000010g^{-3} - 0.7\times 10 ^{-7} g^{-11/3} ,
\end{equation}
and for its imaginary part, we find 
$$
{\rm Im}\; f_2^*(g)\simeq -1.155930g^{1/3} + 0.246594g^{-1/3} -
0.000750g^{-1} -0.001364g^{-5/3}- 
$$
\begin{equation}
-0.000101g^{-7/3} + 0.000003g^{-3} +0.000001g^{-11/3} .
\end{equation}

The approximant (110) possesses the weak--coupling expansion for its
imaginary part, thus, correcting the deficiency of the approximations in
(98). The values of the real and imaginary parts of $f_2^*(g)$ for different 
$g$, as compared to the precise numerical calculations [42,47], are given in
Table III. The maximal percentage error of the real part is $0.7\%$. The
percentage error of the imaginary part cannot be correctly defined, since 
${\rm Im}\; f(g)\rightarrow 0$ as $g\rightarrow 0$. Recall that the
ground--state energy $E(g)$ for Hamiltonian (91) is related, according to
(93), with the function $f(g)$ as 
\begin{equation}
E(g) =\frac{1}{2}f(g) . 
\end{equation}

To write an analytical formula approximating $f(g)$ in the whole range of
the parameter $g\in[0,\infty)$, we may use the asymptotic expansions derived
above. Employing the technique of Sec. II, we can obtain for the real part
the approximants 
\begin{equation}
{\rm Re}\; f_2^*(g,0) =\left [\exp(-6a_1g)+ A_1^6g^2\right ]^{1/6} , 
\end{equation}
\begin{equation}
{\rm Re}\; f_3^*(g,0) =\left [\exp\left\{ -6(a_1+a_2)g\right\} +
A_1^6g^2\right ]^{1/6} + a_2g\left [ 1 +\left (\frac{a_2}{A_2}\right )^{3/2}
g^2\right ]^{2/3} ,
\end{equation}
in which 
$$
a_1=1.5, \qquad a_2 =0.548, \qquad A_1=0.672436, \qquad A_2=0.145202 . 
$$
The behaviour of these approximants is shown in Fig. 2. The maximal
percentage error of $f_2^*(g,0)$ is $24\%$ and that of $f_3^*(g,0)$ 
is $9\%$.

To reach good accuracy for the imaginary part, we need to consider
higher--order approximations. For the imaginary part we find the following
sequence of self--similar approximants: 
\begin{equation}
{\rm Im}\; f_3^*(g,0) = -ag^3\left [\exp\left (-\frac{3b}{4a}g\right ) +
\left (\frac{a}{A}\right )^{3/4}g^2\right ]^{-4/3} ,
\end{equation}
where $a=\frac{3}{4}\sqrt{6},\; b=27\sqrt{6}$, and $A=1.155930$, 
\begin{equation}
{\rm Im}\; f_4^*(g,0) = -ag^3\left\{\left [\exp\left ( -\frac{27b}{28a}g
\right ) +\frac{27}{28}B_1g^2\right ]^{7/6} +\left (\frac{a}{A}\right
)^{9/8} g^3\right\}^{-8/9} ,
\end{equation}
where $a,b$, and $A$ are the same as in (122) and $B_1=0.477045$, 
\begin{equation}
{\rm Im}\; f_5^*(g,0) = -ag^3\left\{\left [\left [\exp\left ( -
\frac{81b}{70a}g\right ) +\frac{81}{70}B_2g^2\right ]^{7/6} +  
\frac{27}{20}B_3g^3\right ]^{10/9} +\left (
\frac{a}{A}\right )^{3/2}g^4 \right\}^{-2/3} ,
\end{equation}
where $B_2=0.178716$ and $B_3=0.496502$, 
$$
{\rm Im}\; f_6^*(g,0) = -ag^3\left\{\left [\left [\left [\exp\left ( -
\frac{243b}{182a}g\right ) +\frac{243}{182}B_4g^2\right ]^{7/6} + 
\frac{81}{52} B_5g^3\right ]^{10/9} +\right.\right. 
$$
\begin{equation}
\left.\left. +\frac{45}{26}B_6g^4\right ]^{13/12}+\left (\frac{a}{A} 
\right )^{15/8}g^5\right\}^{-8/15} ,
\end{equation}
where $B_4=0.073553,\; B_5=0.196401$, and $B_6=0.552922$.

The behaviour of the self--similar approximants (122)--(125) is displayed
in Fig. 3. The curves corresponding to Eqs. (124) and (125) almost
coincide in this picture. Fig. 3 clearly demonstrates the convergence of
the sequence $\{ {\rm Im}\;f_k^*(g,0)\}$ to exact data marked by diamonds.

\section{Nonlinear Hamiltonians}

Here we show that our approach is applicable not only to linear problems of
quantum mechanics but to nonlinear problems as well. 

\vspace{3mm}

{\bf A. One--Dimensional Case}

\vspace{1mm}

Consider the nonlinear Hamiltonian 
\begin{equation}
H =-\frac{1}{2}\frac{d^2}{dx^2} +\frac{1}{2}x^2 +g\psi^2(x) ,
\end{equation}
in which $x\in(-\infty,+\infty), \; g\in(-\infty,+\infty)$, and $\psi(x)$ is
a wave function.

The Hamiltonian (126) is a prototype of the $\varphi^4$ model of quantum
field theory. There exists a controversy in the interpretation of the
so--called "triviality" of this theory (see discussion in Refs. [48-50]).
Therefore, developing methods that could successfully deal with nonlinear
Hamiltonians of the type (126) could be useful for $\varphi^4$ quantum field
theories, as well as for those field theories that include scalar--field
terms, like (126), in their Lagrangians, e.g., as in the Higgs model [51].
Another important application of the nonlinear Hamiltonian of the form (126)
is for describing properties of atoms confined in magnetic traps [52-54].
Such magnetically trapped atoms of $^{87}Rb,\; ^{23}Na$, and $^7Li$, as has
been observed recently [55-57], can exhibit the phenomenon of Bose--Einstein
condensation. The possibility of the direct observation of Bose condensation
distinguishes these alkali gases from liquid $^4He$ where this condensation
could be investigated only indirectly (see the related discussion in Refs.
[58-63]). The system of condensed trapped atoms corresponds to the ground
state of the Hamiltonian (126). We consider here the one--dimensional case
which serves as an illustration of the applicability of the method.

It is worth emphasizing that the coupling parameter $g$ in Eq. (126), when a
system of $N$ condensed atoms is considered, is proportional to $N$. Because
of this, for $N\gg 1$, the coupling parameter can become large and,
consequently, perturbation theory in powers of $g$ is of no sense. Although
some thermodynamic characteristics of trapped atoms can be approximately
analyzed disregarding their interactions, that is, nonlinearity [64-67], but
a correct description certainly has to take into account these interactions
[68], i.e., nonlinearity, since for the cases related to experiment [55-57]
the effective coupling is strong, $g\gg 1$.

To derive analytical formulas for the spectrum of the Hamiltonian (126), we
need to find appropriate asymptotic expansions. Let us start with the
Hamiltonian 
\begin{equation}
H_0 =-\frac{1}{2}\frac{d^2}{dx^2} +\frac{u^2}{2} x^ 2 ,
\end{equation}
in which $u$ is a trial parameter. Employing perturbation theory with
respect to the perturbation 
$$
\Delta H=\frac{1}{2}\left ( 1 -u^2\right ) x^2 +g\psi^ 2(x) , 
$$
we may find the $k$--order approximation 
\begin{equation}
E_k(g,u) =\left ( n+\frac{1}{2}\right ) F_k(g,u)
\end{equation}
for the energy levels labelled by $n=0,1,2,\ldots$. For the function $F_k$
defined in (128) we have 
\begin{equation}
F_0(g,u) = u , \qquad
F_1(g,u)=\frac{1}{2}\left ( u +\frac{1}{u}\right ) +
\frac{J_n}{n+1/2}g\sqrt{u} ,
\end{equation}
where 
$$
J_n=\frac{1}{\pi 2^nn!}\int_{-\infty}^{+\infty} \exp (-2x^2) H_n^4(x)dx ; 
$$
$H_n(x)$ being a Hermite polynomial. In particular, 
$$
J_0=\frac{1}{\sqrt{2\pi}} , \qquad J_1 =\frac{3}{4\sqrt{2\pi}} , \qquad 
J_2= \frac{41}{64\sqrt{2\pi}} . 
$$

The control function $u=u(g)$ can be found from the fixed--point condition 
\begin{equation}
\frac{\partial}{\partial u} F_1(g,u) = 0 .
\end{equation}
The latter, with the notation 
\begin{equation}
\alpha\equiv\frac{J_n}{n+1/2} g ,
\end{equation}
yields 
\begin{equation}
u^2 +u^{3/2}\alpha - 1 = 0 .
\end{equation}
Substituting the solution $u(g)$ to Eq. (132) into (129), we get 
\begin{equation}
f_k(g) = F_k(g,u(g)) .
\end{equation}
For instance, 
\begin{equation}
f_1(g) =\frac{1}{2}\left ( \frac{3}{u} - u\right ) , \qquad u=u(g) .
\end{equation}

In the weak--coupling limit, $\alpha\rightarrow 0$ when $g\rightarrow 0$,
according to (131). Then (134) gives 
\begin{equation}
f_1(g) \simeq 1 +\alpha -\frac{1}{8}\alpha^2 +\frac{1}{32}\alpha^3 - 
\frac{1}{128}\alpha^4 + \frac{3}{2048}\alpha^5 .
\end{equation}

In the strong--coupling limit, when $g\rightarrow\infty$ together with 
$\alpha\rightarrow\infty$, Eq. (134) leads to 
\begin{equation}
f_1(g) \simeq \frac{3}{2}\alpha^{2/3} +\frac{1}{2}\alpha^{-2/3} -
\frac{1}{6} \alpha^{-2} +\frac{7}{54}\alpha^{-10/3} .
\end{equation}

In the case of a negative coupling parameter $g< 0$, the weak--coupling
limit, when $g\rightarrow -0$ and $\alpha\rightarrow -0$, gives the same
asymptotic expansion as (135). However, the strong--coupling limit, when 
$g\rightarrow -\infty$ and $\alpha\rightarrow -\infty$, is different from
(136). If $\alpha\rightarrow -\infty$, then Eq. (134) behaves as 
\begin{equation}
f_1(g)\simeq -\frac{\alpha^2}{2} +\frac{1}{2\alpha^2} -\frac{1}{2\alpha^6} + 
\frac{3}{2\alpha^{10}} -\frac{13}{2\alpha^{14}} .
\end{equation}

In the region of negative $g < 0$, the energy (128) is positive for small 
$|g|$, and, as $g$ diminishes, the energy becomes zero at a critical value 
$g_c$. The latter can be found from the definition 
\begin{equation}
f_1(g_c) = 0 .
\end{equation}
The form (134) shows that equality (138) holds true for $u_c^2=3$. Then, Eq.
(132) immediately gives 
\begin{equation}
\alpha_c=-\frac{2}{3^{3/4}} = -0.87738 .
\end{equation}
Because of the relation (131), one has 
\begin{equation}
g_c=-\left ( n+\frac{1}{2}\right )\frac{\alpha_c}{J_n} .
\end{equation}
For example, for the ground--state level, with $n=0$, one finds 
\begin{equation}
g_c=-1.09964 \qquad (n=0).
\end{equation}
For $g< g_c$, the energy becomes negative, which implies the instability 
of the system.

Note that invoking the notation (131), we have managed to write the
asymptotic expansions (135), (136), and (137) for arbitrary energy levels.
Using the derived expansions, we can construct analytical formulas for the
whole spectrum of the considered nonlinear Hamiltonian. Thus, following the
standard way of Sec. II, we find for $\alpha\geq 0$ the approximant 
\begin{equation}
f_2^*(\alpha,0) =\left [\left (1 +a_1A_1^3\sqrt{8A_2^3}\alpha\right )^{2/3}
+ A_1^3\alpha^2\right ]^{1/3} ,
\end{equation}
where $a_1=1,\; A_1=3/2$, and $A_2=1/2$. Similarly, we can write down 
\begin{equation}
f_2^*(\alpha,\infty) =\left [ 1 + A_1^4\alpha\left (\frac{16A_2}{5A_1}\tau_2
+\alpha^{4/3}\right )^{5/4}\right ]^{1/4} ,
\end{equation}
where $A_1$ and $A_2$ are the same as before and 
$$
\tau_2=\frac{5}{A_2}\left (\frac{a_1^4}{2^{12}A_1^{11}}\right )^{1/5} =
0.776 . 
$$
For $\alpha\leq 0$, we may construct 
\begin{equation}
f_2^*(\alpha,-0) = 1+a_1\alpha\left [\exp\left ( - \frac{2a_2}{a_1}
\alpha\right ) +\frac{B_1^2}{a_1^2}\alpha^2\right ]^{1/2} ,
\end{equation}
with $a_2=-1/8$ and $B_1=-1/2$. Another approximant is 
\begin{equation}
f_3^*(\alpha,-\infty) = B_1\alpha^2 + B_2\left [ B_2^4 +|\alpha|\left (
\alpha^4 -\frac{16B_3}{7B_2}\tau_3\right )^{7/4}\right ]^{-1/4} ,
\end{equation}
where $B_2=1/2,\; B_3=-1/2$, and 
$$
\tau_3=\frac{7}{|B_3|}\left (\frac{a_1^4B_2^{23}}{2^{20}}\right )^{1/7} =
0.198 . 
$$

The behaviour of approximants (142)--(145) is such that $f_2^*(\alpha,0)$ 
practically coincides with $f_2^*(\alpha,\infty)$ for $\alpha\geq 0$, while 
$f_2^*(\alpha,-0)$ coincides with $f_3^*(\alpha,-\infty)$ for $\alpha\leq 0$. 
A slight difference between these approximants is noticeable only in the
region $-1\leq\alpha\leq 1$, as is shown in Fig. 4.

\vspace{3mm}

{\bf B. Radial Model}

In the previous section we considered a one--dimensional nonlinear problem.
It is straightforward to apply the same approach to nonlinear problems of
higher dimensionalities. As an illustration, we consider a spherically
symmetric model with the nonlinear radial Hamiltonian 
\begin{equation}
H =-\frac{1}{2}\frac{d^2}{dr^2} +\frac{l(l+1)}{2r^2} +\frac{1}{2}r^2 +
gR_{nl}^2(r) ,
\end{equation}
in which $r\in[0,\infty), \; n=0,1,2,\ldots, \; l=0,1,2,\ldots$, the
coupling parameter $g\in(-\infty,+\infty)$, and $R_{nl}$ is a radial wave
function.

Starting from the trial Hamiltonian 
\begin{equation}
H_0 =-\frac{1}{2}\frac{d^2}{dr^2} +\frac{l(l+1)}{2r^2} +\frac{u^2}{2}r^2 ,
\end{equation}
with a trial parameter $u$, we invoke perturbation theory with respect to
the perturbation $\Delta H=H-H_0$. For the $k$--order approximation of the
spectrum we may write 
\begin{equation}
E_k(g,u) \equiv\left ( 2n + l +\frac{3}{2}\right ) F_k(g,u) .
\end{equation}
Introduce the effective coupling parameter 
\begin{equation}
\alpha\equiv\frac{J_{nl}}{2n+l+3/2} g ,
\end{equation}
in which 
$$
J_{nl}=\left [\frac{2n!}{\Gamma(n+l+3/2)}\right ]^2\int_0^\infty
r^{4(l+1)}e^{-2r^2}\left [ L_n^{l+1/2}(r^2)\right ]^4 dr , 
$$
where $\Gamma$ is a gamma-function and $L_n^l$ is an associated Laguerre
polynomial. For the particular cases that will be analysed in what follows,
we have 
$$
J_{00}=\frac{3}{2\sqrt{2\pi}} , \qquad 
J_{01} =\frac{35}{24\sqrt{2\pi}} , \qquad
J_{10}=\frac{147}{128\sqrt{2\pi}}, \qquad 
J_{02}=\frac{231}{160\sqrt{2\pi}} . 
$$

Accomplishing the same steps as in the previous section, we can find the
spectrum 
\begin{equation}
e_{nl}(g) =\left ( 2n +l +\frac{3}{2}\right ) f_1(g) ,
\end{equation}
in which $f_1(g)$ is defined identically to (133). The control function $u(g)
$ is given by the solution of equation (132), only with $\alpha$ introduced
in (149). With this renotation for the parameter $\alpha$, all asymptotic
expansions for $f_1(g)$ are the same as in Sec. VII. Therefore, the
corresponding approximants $f_k^*$ will have the same forms (142)-(145).

What differs the considered nonlinear radial model from the one--dimensional
nonlinear case is that the energy levels are labelled now by two quantum
indices, $n$ and $l$. For some combinations of these indices, the specific
effect of {\it level crossing}, when varying $g$, may occur. To illustrate
this, let us analyse several lower levels of the spectrum (150). In the
weak--coupling limit we may find 
$$
e_{00}\simeq \frac{3}{2}\left ( 1 +\frac{g}{\sqrt{2\pi}}\right ) , \qquad
e_{01}\simeq \frac{5}{2}\left ( 1 +\frac{7g}{12\sqrt{2\pi}}\right ) , 
$$
$$
e_{10}\simeq \frac{7}{2}\left ( 1 +\frac{21g}{64\sqrt{2\pi}}\right ) ,
\qquad e_{02}\simeq \frac{7}{2}\left ( 1 +\frac{33g}{80\sqrt{2\pi}}
\right ) , \qquad g\rightarrow 0 . 
$$
While for the strong--coupling limit, we have 
$$
e_{00}(g)\simeq 1.219334 g^{2/3} , \qquad e_{10}(g)\simeq 1.353502 g^{2/3} 
$$
$$
e_{01}(g)\simeq 1.418783 g^{2/3} , \qquad e_{02}(g)\simeq 1.576577 g^{2/3} ,
\qquad g\rightarrow \infty. 
$$
As is seen, at small $g$ the energy level $e_{01}$ is lower than $e_{10}$,
but at large $g$, vice versa, the level $e_{10}$ becomes lower than $e_{01}$.

Another difference with the one--dimensional nonlinear case is the value of
the critical coupling constant $g_c$ at which the energy level $e_{nl}(g)$
crosses zero. This critical parameter is defined as in (138) and gives the
same $\alpha_c$ as in (139). But, because of the different relation (149)
between $\alpha$ and $g$, we now get, instead of (140), the critical
coupling 
$$
g_c= -\left ( 2n+l+\frac{3}{2}\right )\frac{\alpha_c}{J_{nl}} . 
$$
For the ground state--level, instead of (141), we find 
$$
g_c=-2.19927 \qquad (n=l=0), 
$$
so that $|g_c|$ is about twice larger than that for the one--dimensional
case.

\section{Wave Functions}

In this section we show that the approach we have developed permits us to
construct analytical expressions not only for energy levels but for wave
functions as well. We shall concentrate our attention on the most
interesting, from our point of view, case of nonlinear equations.

\vspace{3mm}

{\bf A. Nonlinear Schr\"odinger Equation}

\vspace{1mm}

Consider the equation 
\begin{equation}
-\frac{1}{2}\frac{d^2\psi}{dx^2} +\frac{1}{2}x^2\psi + g\psi^3 = E\psi ,
\end{equation}
in which $x\in(-\infty,+\infty)$ and $g\in[0,\infty)$. The nonlinear
Schr\"odinger equations of this type are used for describing Bose--condensed
atoms in magnetic traps [55-57]. This kind of equations is also often called
the Gross--Ginzburg--Pitaevskii equation [69-72]. Condensed atoms correspond
to the ground state of (151), which is assumed in what follows.

The wave function $\psi=\psi(x)$ is normalized by the condition 
\begin{equation}
\int_{-\infty}^{+\infty} |\psi(x)|^2 dx = 1 .
\end{equation}
From Eq. (151) it follows that the wave function is an even function, so
that $\psi(x)=\psi(-x)$. Because of this, it has an expansion 
\begin{equation}
\psi(x)\simeq c_0 + c_2x^2 +c_4x^4 ,
\end{equation}
as $x\rightarrow 0$, in even powers of $x$. For large $x$, the harmonic term
in (151) becomes predominant, and the wave function has the asymptotic
behaviour 
\begin{equation}
\psi(x)\simeq A\exp\left (-\frac{1}{2}x^2\right )
\end{equation}
as $x\rightarrow\infty$. Our aim is to find an analytical expression for 
$\psi(x)$ valid in the whole region of $x\in(-\infty,+\infty)$. Note that
Pad\'e approximants are not able to interpolate between such different types
of behaviour as the power law in (153) and exponential in (154). But using
the self--similar interpolation of Sec. II, we easily obtain the
self--similar approximant 
\begin{equation}
\psi^*(x) = A\exp\left\{ -\frac{1}{2}x^2 + ax^2\exp(-bx^2)\right\} ,
\end{equation}
where $\psi^*(x)\equiv \psi_3^*(x,0)$ and 
$$
A=c_0, \qquad a=\frac{1}{2} +\frac{c_2}{c_0}, \qquad b=-\frac{c_4}{ac_0^2} . 
$$
Expression (155) acquires a transparent physical meaning when written in the
form 
$$
\psi^*(x) = A\exp\left\{ -\frac{x^2}{\xi^2(x)}\right\} , 
$$
in which 
$$
\xi(x) =\left (\frac{1}{2} - a\exp\left\{ -bx^2\right\}\right )^{-1/2} 
$$
plays the role of an effective correlation length.

The parameters $a,\; b$, and $A$ in (155) are not independent. The relation
between them can be found if we expand (155) in powers of $x$ and substitute
this expansion into Eq. (151). For small $x$, Eq. (155) gives 
$$
\psi^*(x)\simeq c_0 +c_2x^2 +\bar c_4 x^4 \qquad (x\rightarrow 0) , 
$$
where $\bar c_4=c_4 +c_2^2/2c_0$. Substituting this into (151) and equating
the terms at like powers of $x$, we find the relations 
\begin{equation}
c_2=c_0\left ( gc_0^2 -E\right ) , \qquad c_4 =\frac{c_0}{12}\left ( 1
+4gc_0^2E - 4E^2\right ) .
\end{equation}
The latter yield the equalities 
\begin{equation}
a =\frac{1}{2} + gA^2 - E , \qquad b=\frac{2(1-2a)E-1}{12aA}
\end{equation}
showing that among four parameters, $a,\; b,\; A$, and $E$, there are only
two independent. Two additional equations for defining all parameters are
the normalization condition (152) and the definition of the energy 
\begin{equation}
E^*(g) =\left (\psi^*,H\psi^*\right ) ,
\end{equation}
where the Hamiltonian $H$ is the same as in (126) and the notation 
$E^*(g)\equiv E$ stresses that the energy is obtained by using the
self--similar approximant (155).

The values of $a,\; b$ and $E^*$ depend on the coupling parameter $g$. Thus,
for the weak--coupling limit $g\rightarrow 0$ we have 
$$
E^*(g)\simeq \frac{1}{2} +\frac{1}{\sqrt{2\pi}}g , \qquad A^2 =c_0^2 \simeq 
\frac{1}{\sqrt{\pi}} . 
$$
Then, relations (156) give 
$$
\frac{c_2}{c_0}\simeq -\frac{1}{2} +\frac{\sqrt{2}-1}{\sqrt{2\pi}}g , \qquad 
\frac{c_4}{c_0} \simeq -\frac{\sqrt{2}-1}{b\sqrt{\pi}} g . 
$$
Respectively, from (157) we get 
$$
a\simeq\frac{\sqrt{2}-1}{\sqrt{2\pi}} g=0.165247 g, \qquad 
b\simeq\frac{1}{3\sqrt{2}} =0.235702 . 
$$
This demonstrates that the function (155) reduces to the Gaussian form when 
$g\rightarrow 0$.

The variational Gaussian function 
\begin{equation}
\psi_G(x) =\left (\frac{u}{\pi}\right )^{1/4}\exp\left (-\frac{u}{2}x^2
\right )
\end{equation}
is often used not only for small $g\ll 1$ but for arbitrary 
$g\in[0,\infty)$, with the effective frequency $u=u(g)$ defined by the 
minimum of the energy $(\psi_G,H\psi_G)$, which gives 
$$
u^2 + u^{3/2}\sqrt{\frac{2}{\pi}}\; g -1 = 0 . 
$$
Such a variational energy is very close, with a deviation not more than
several percent, to the energy 
$$
E_2^*(g) =\frac{1}{2}f_2^*(\alpha,0) , \qquad 
\alpha\equiv\sqrt{\frac{2}{\pi}}\; g , 
$$
corresponding to Eq. (142), which results in 
\begin{equation}
E_2^*(g) =\frac{1}{2}\left [\left ( 1 + \frac{27}{4\sqrt{2\pi}}g
\right )^{2/3} +\frac{27}{4\pi}g^2\right ]^{1/3} .
\end{equation}

Another very often used approximation for treating Bose--condensed atoms in
harmonic traps is the Thomas--Fermi approximation, (see e.g. [73-75]) which
for Eq. (151) leads to the wave function 
$$
\psi_{TF}(x) =\left (\frac{x_0^2 -x^2}{2g}\right )^{1/2} , \qquad 
|x|\leq x_0 , 
$$
\begin{equation}
\psi_{TF}(x)=0 , \qquad |x|\geq x_0 ,
\end{equation}
in which 
$$
x_0\equiv\sqrt{2 E_{TF}} =\left (\frac{3}{2}g\right )^{1/3} . 
$$
The energy in the Thomas--Fermi approximation is obtained from the
normalization condition (152) giving 
\begin{equation}
E_{TF}(g) =\frac{1}{2}\left (\frac{3}{2}g\right )^{2/3} .
\end{equation}
The Thomas--Fermi approximation is assumed to be valid for large 
$g\rightarrow\infty$. However, even then the wave function (161) is correct
only for $x\ll x_0$, where it has an expansion 
$$
\psi_{TF}(x)\simeq c_0 + c_2x^2 + c_4x^ 4 
$$
with the coefficients 
$$
c_0=\frac{x_0}{\sqrt{2g}} , \qquad c_2=-\frac{c_0}{2x_0^2} , \qquad 
c_4=- \frac{c_0}{8x_0^4} . 
$$
The behaviour of the function (161) near the boundary $x=x_0$ is not
correct. Also, this function is not appropriate to evaluate the mean kinetic
energy, producing a divergence for any $g$ (see discussion in [73,75]). In
order to understand when the total energy (162) for the Thomas--Fermi
approximation starts giving reasonable results, we present in Fig. 5 the
energies (158), (160), and (162). The first two energies, $E^*(g)$ and 
$E_2^*(g)$, almost coincide with each other, having correct asymptotic
behaviour in the weak as well as in the strong coupling limits. The
Thomas--Fermi energy $E_{TF}(g)$ possesses an incorrect weak--coupling limit
and becomes a reasonable approximation starting from $g=7$.

The density 
\begin{equation}
n(x) =|\psi(x)|^2
\end{equation}
for the corresponding wave functions and $g=0.2$ is presented in Fig. 6,
where the density $n^*(x)=|\psi^*(x)|^2$ of the self--similar approximation 
(155) practically coincides with the density $n_G(x)=|\psi_G(x)|^2$ of the 
Gaussian approximation (159), as it should be in the weak--coupling limit.
In this limit, the behaviour of the density $n_{TF}(x)=|\psi_{TF}(x)|^2$
of the Thomas--Fermi approximation is not correct. As is known [73,75],
the latter approximation is incorrect near the boundary even in the
strong--coupling case. Then the density $n(x)$ in the self--similar
approximation is close, except near the boundary, to that of the
Thomas--Fermi approximation. The self--similar approximation $n^*(x)$
coincides with $n_{TF}(x)$ for small $x$ and smoothes the incorrect
behaviour of $n_{TF}(x)$ around the boundary. In the strong--coupling
limit, the density $n_G(x)$ of the Gaussian approximation is not accurate.

The direct evaluation of the accuracy of each approximation can be done by
calculating the residual term 
\begin{equation}
R(x) \equiv H\psi(x) -(\psi,H\psi)
\end{equation}
for Eq. (151), where $H$ is defined in (126) and $\psi(x)$ is a wave
function of the corresponding approximation. The residual terms for 
$g\gg 1$ for the self--similar approximation (155) is practically zero,
meaning that (155) is an almost exact solution of Eq. (151). For the
Gaussian approximation (159), the residual term is much larger, telling that
this approximation is much less accurate. And the residual for the
Thomas--Fermi approximation is divergent at the boundary point $x_0$,
though far from this point it is close to zero.

The integral characteristic of accuracy of the corresponding solutions is
the dispersion 
\begin{equation}
\sigma(\psi) \equiv \left [ \int_{-\infty}^{+\infty} |R(x)|^2 dx
\right ]^{1/2} .
\end{equation}
We calculated this quantity for $0\leq g\leq 100$. The maximal dispersion
for the self--similar approximation (155) is around $1$, for the Gaussian
approximation (159), it is about $20$, and for the Thomas--Fermi
approximation, it is divergent.

In this way, the self--similar wave function (155) is the most accurate
solution of the nonlinear Schr\"odinger equation (151), as compared to the
Gaussian and Thomas--Fermi approximations. This function (155) represents
the exact solution very well for all $x$ and $g$. In the weak--coupling
limit $g\rightarrow 0$, it becomes close to the Gaussian form, and in the
strong--coupling limit, it approaches the Thomas--Fermi wave function for
all $x$ except the boundary where it smoothes the incorrect behaviour of the
latter function. The crossover point between the weak--coupling and
strong--coupling regimes occurs, as numerical calculations show, at around 
$g\approx 5$. This crossover point can also be evaluated, by an order of
magnitude, analytically as follows. Notice that the characteristic length
for the Gaussian function (159) is $x_G=\sqrt{2/u}$ with $u\approx 1$, and
that such a length for the Thomas--Fermi function (161) is $x_0=(3g/2)^{1/3}$. These characteristic lengths, typical of the weak--coupling and
strong--coupling regimes, respectively, coincide, that is $x_G=x_0$, at 
$g\approx 2^{5/2}/3\approx 2$.

\vspace{3mm}

{\bf B. Vortex Filament Equation}

\vspace{1mm}

Now we shall show that our approach permits us to find accurate analytical
approximations for the function describing the structure of vortex
filaments. Considering an unbounded Bose system and making in the nonlinear
Schr\"odinger equation the substitution $\psi(\stackrel{\rightarrow}{r}) =
f(r)e^{i\varphi}$, in which $r$ and $\varphi$ are dimensionless polar
coordinates, one comes [70,72] to the equation 
\begin{equation}
\frac{d^2f}{dr^2} +\frac{1}{r}\frac{df}{dr} -\frac{f}{r^2} + f - f^3 = 0 .
\end{equation}
The solution to this equation is usually obtained numerically [70,76,77].
Here we shall construct a sequence of analytical approximations for the
solution to Eq. (166) and compare them with the known numerical data. Note
that the equations similar to (166) have been considered as well for
describing magnetic solitons [78], isomeric states of quantum fields [79],
and vortices of complex scalar fields [80]. Therefore, the possibility of
deriving accurate analytical solutions to these equations is important for
many applications, such as condensed Bose gas, superfluid helium, magnets in
strong magnetic fields, and different models of quantum fields.

At small $r\rightarrow 0$, the solution to Eq. (166) has the asymptotic
expansion 
\begin{equation}
f(r)\simeq cr\left ( 1 -\frac{1}{8}r^2\right ) ,
\end{equation}
where $c$ is a constant. At large $r\rightarrow\infty$, one gets 
\begin{equation}
f(r)\simeq 1 -\frac{1}{2}r^{-2} -\frac{9}{8}r^{-4} -\frac{169}{16}r^{-6} .
\end{equation}
Employing the approach of Sec. II, we easily obtain the following sequence
of self--similar approximants: 
$$
f_2^*(r,0) = c_2r\left ( 1 +\frac{1}{4}r^2 \right )^{-1/2} , 
$$
$$
f_3^*(r,0) = c_3r\left ( 1+\frac{1}{2}r^2 +\frac{1}{4}r^4 \right )^{-1/4} , 
$$
$$
f_4^*(r,0) = c_4r\left ( 1 +\frac{3}{4}r^2 +\frac{3}{16}r^4 + 
\frac{1}{16} r^6 \right )^{-1/6} , 
$$
\begin{equation}
f_5^*(r,0) = c_5r\left ( 1 + r^2 +\frac{9}{70}r^4 +\frac{1}{35}r^6 + 
\frac{1}{140}r^8 \right )^{-1/8} ,
\end{equation}
in which the coefficients, defined so that to give the correct asymptotic
expansions, are 
\begin{equation}
c_2=4^{-1/2} =0.5 , \qquad c_3= 4^{-1/4} = 0.707 , \qquad
c_4=16^{-1/6} =0.630 , \qquad c_5=140^{-1/8} = 0.539 .
\end{equation}
The behaviour of the approximants $f_k^*(r,0)$ is shown in Fig. 7, compared
with numerical data [70,76,77]. As it can be concluded from this figure, 
$f_5^*(r,0)$ is a very accurate solution.

\section{Conclusion}

We have developed an approach for obtaining analytical solutions of
quantum--mechanical problems. This approach makes it possible, starting from
asymptotic expansions having sense only in the vicinity of limiting points,
to derive interpolation formulas valid in the whole range of variables. The
developed method is rather general and can be applied to various problems.
We demonstrated its applicability to several quantum--mechanical models,
such as different anharmonic oscillators, double--well potentials, resonance
models with quasistationary states, and nonlinear Hamiltonians. The method
permits one to construct accurate analytical expressions for energy levels
as well as for wave functions. It is important that this method provides a
regular procedure for deriving a convergent sequence of subsequent
approximations, so that it is possible to reach the desired accuracy by
calculating higher--order approximations. The idea of the approach is based
on the self--similar approximation theory [8-17], this is why we call the
method developed in the present paper the self--similar interpolation. The
method can find numerous practical applications, for example, for analysing
spectral properties of atoms and molecules, for studying the physics of
quantum dots, and for investigating the behaviour of Bose condensed gases in
magnetic traps.

\vspace{5mm}

{\bf Acknowledgement}

\vspace{2mm}

We appreciate financial support from the National Science and Technology
Development Council of Brazil and from the University of Western Ontario,
Canada.

\newpage

\begin{table}

\caption{Percentage errors of the self--similar approximants for the
ground--state energy of the sextic oscillator, as compared to numerical 
data for $e(g)$.}

\vspace{5mm}

\begin{tabular}{|c|c|c|c|c|c|c|c|c|c|} \hline
$g$ & $e(g)$ & $\varepsilon_1^*(g,0)$ & $\varepsilon_1^*(g,\infty)$ & 
$\varepsilon_1^*(g)$ & $\varepsilon_2^*(g,0)$ & $\varepsilon_3^*(g,0)$ & 
$\varepsilon_4^*(g,0)$ & $\varepsilon_5^*(g,0)$ & $\varepsilon_6^*(g,0)$ \\ 
\hline
0.1 & 0.586945 & $-$8.25 & 7.25 & $-$0.50 & $-$5.45 & $-$3.99 & $-$3.06 & $-$
2.46 & $-$1.95 \\ 
0.5 & 0.717813 & $-$10.56 & 5.88 & $-$2.34 & $-$4.71 & $-$2.48 & $-$1.44 & $-$
0.88 & $-$0.53 \\ 
2 & 0.915219 & $-$8.47 & 3.99 & $-$2.24 & $-$2.40 & $-$0.82 & $-$0.33 & $-$
0.11 & $-$0.06 \\ 
50 & 1.858487 & $-$2.47 & 1.03 & $-$0.67 & $-$0.16 & $-$0.03 & 0 & 0 & 0 \\ 
1000 & 3.850896 & $-$0.59 & 0.25 & $-$0.16 & $-$0.01 & 0 & 0 & 0 & 0 \\ 
\hline
\end{tabular}

\newpage

\caption{Percentage errors of the self--similar approximants for the
ground--state energy of the octic oscillator.} 

\vspace{5mm}

\begin{tabular}{|c|c|c|c|c|c|c|c|c|c|} \hline
$g$ & $e(g)$ & $\varepsilon_1^*(g,0)$ & $\varepsilon_1^*(g,\infty)$ & 
$\varepsilon_1^*(g)$ & $\varepsilon_2^*(g,0)$ & $\varepsilon_3^*(g,0)$ & 
$\varepsilon_4^*(g,0)$ & $\varepsilon_5^*(g,0)$ & $\varepsilon_6^*(g,0)$ \\ 
\hline
0.1 & 0.620514 & $-$12.01 & 7.82 & $-$2.10 & $-$7.90 & $-$5.72 & $-$4.27 & 
$-$3.30 & $-$2.66 \\ 
0.5 & 0.745510 & $-$12.54 & 6.10 & $-$3.22 & $-$5.77 & $-$3.15 & $-$1.88 & 
$-$1.21 & $-$0.81 \\ 
2 & 0.911090 & $-$9.66 & 4.39 & $-$2.64 & $-$3.13 & $-$1.26 & $-$0.60 & 
$-$0.27 & $-$0.17 \\ 
50 & 1.594327 & $-$3.38 & 1.52 & $-$0.93 & $-$0.43 & $-$0.08 & $-$0.02 & 0 & 
0 \\ 
1000 & 2.833102 & $-$1.06 & 0.49 & $-$0.28 & $-$0.04 & 0 & 0 & 0 & 0 \\ 
\hline
\end{tabular}

\vspace{3cm}

\caption{The double energy $f_2^*(g)=2E^*(g)$ of the lowest 
quasistationary state, as compared to numerical values $f(g)$.}

\vspace{5mm}

\begin{tabular}{|c|c|c|c|c|} \hline
$g$ & Re$f(g)$ & Re$f_2^*(g)$ & $-$Im$f(g)$ & $-$Im$f_2^*(g)$ \\ \hline
0.01 & 0.984428 & 0.984429 & 0.000000 & 0.000003 \\ 
0.02 & 0.967451 & 0.967477 & 0.000001 & 0.000035 \\ 
0.05 & 0.900673 & 0.901116 & 0.006693 & 0.004888 \\ 
0.1 & 0.794881 & 0.791404 & 0.089412 & 0.090883 \\ 
0.2 & 0.72882 & 0.728985 & 0.27735 & 0.281642 \\ 
0.5 & 0.7477 & 0.751513 & 0.6100 & 0.613077 \\ 
1.0 & 0.8297 & 0.834876 & 0.9097 & 0.911547 \\ 
2.0 & 0.964 & 0.971001 & 1.260 & 1.261483 \\ 
5.0 & 1.23 & 1.238149 & 1.84 & 1.832649 \\ \hline
\end{tabular}

\end{table}

\newpage

\begin{center}
{\bf Figure Captions}
\end{center}

{\bf Fig. 1}. The free energy of the double--well model given by the
approximants $f_1(g)$ (dashed line), $f_2(g)$ (short--dashed line),
and $f_2^*(g)$ (solid line). Crosses correspond to the exact values of
function (61).

\vspace{5mm}

{\bf Fig. 2}. The real part of the double energy for the lowest
quasiresonance state represented by the approximants Re$f_2^*(g,0)$ (dashed
line) and Re$f_3^*(g,0)$ (solid line), compared to exact values marked by
diamonds.

\vspace{5mm}

{\bf Fig. 3}. The modulus of the imaginary part of the double energy for 
the lowest quasiresonance state given by the approximants Im$f_3^*(g,0)$
(short--dashed line), Im$f_4^*(g,0)$ (dashed line), and
Im$f_6^*(g,0)$ (solid line). Exact data are shown by diamonds.

\vspace{5mm}

{\bf Fig. 4}. The self--similar approximants $f_2^*(\alpha,0)$ (solid line)
and $f_2^*(\alpha,\infty)$ (dashed line) for $\alpha\geq 0$ and the
approximants $f_2^*(\alpha,-0)$ (solid line) and $f_3^*(\alpha,-\infty)$
(dashed line) for $\alpha\leq 0$, in the region $-1\leq\alpha\leq 1$.

\vspace{5mm}

{\bf Fig. 5}. The ground--state energy for the nonlinear Schr\"odinger
equation (151) for the self--similar approximants $E^*(g)$ (solid line) and 
$E_2^*(g)$ (dashed line) and for the Thomas--Fermi approximation $E_{TF}(g)$
(short--dashed line).

\vspace{5mm}

{\bf Fig. 6}. The density (163) for the corresponding wave functions in the
self--similar approximation (155) (solid line), Gaussian approximation (159)
(short--dashed line), and Thomas--Fermi approximation (dashed line) for
$g=0.2$.

\vspace {5mm}

{\bf Fig. 7} The self--similar approximants $f_k^*(r,0)$, defined in Eq.
(169), describing the structure of a vortex filament. The solid line is for 
$f_2^*(r,0)$, long--dashed line is for $f_3^*(r,0)$, short--dashed line is
for $f_4^*(r,0)$, and the dotted line is for $f_5^*(r,0)$. Diamonds
represent exact numerical data.

\end{document}